# Extracting Super-resolution Structures inside a Single Molecule or Overlapped Molecules from One Blurred Image

Yaohua Xie[*] (http://orcid.org/0000-0001-6780-3156)
Yaohua.Xie@hotmail.com, or Edward.Y.Sheffield@hotmail.com

**Abstract:** In some super-resolution techniques, adjacent points are illuminated at different times. Thereby, their locations and light intensities can be detected even if the images are very blurred due to diffraction. According to conventional theories, the points' inner details cannot be recovered because the images' high frequency components are removed due to the diffraction-limit. But this study finds an exception, and full information can be extracted from a diffraction-blurred image. In such a "resolvable condition", neither profile nor detail information is damaged by diffraction. Thereby, it can be recovered reversibly by solving equation systems in spatial domain or frequency domain. This condition is tightly relevant to the imaging condition of existing super-resolution techniques. Based on the condition, a method is proposed which can achieve unlimited high resolutions in principle, and its effectiveness is demonstrated by both theoretical analysis and simulation experiments. It can also work without any observed image outside the region of interest. Simulation experiments also show its tolerance to certain level of noise.

## 1. Introduction

When an object (sample) is imaged by a conventional light microscope, the result is not an ideal image which shows sharp details. Instead, it is equivalent to the ideal image convolved with a Point Spread Function (PSF) whose central part is called Airy disk. Therefore, even a point is infinitely small, its image is an Airy-disk-shaped pattern rather than an ideal point. In 1873, Ernst Abbe first described the diffraction-limit: for two points with a distance less than a half of visible light's wavelength, i.e., about 200~300nm, their images overlap each other and cannot be resolved. Usually, the Rayleigh Criterion can be adopted to judge whether points are resolvable. Samples' structures smaller than the diffraction-limit were not resolvable with such microscopes until super-resolution techniques emerged. These techniques are mainly divided into two categories [1]. The first category uses structural-illumination to image the sample multiple times, and then processes the resulting images to get a super-resolution image. Representative techniques: STED [2], RESOLFT [3], SIM [4], NL-SIM [5], et al. The second category manages to turn on individual molecules at different times, i.e., separates them by time, and then also reconstructs a super-resolution image. Representative techniques: PALM [6], STORM [7,] PAINT [8], et al. Besides, a technique named MINFLUX [9] combines the advantages of the two categories. It can localize individual molecules with ultra-high precision. There is a different technique named Expansion Microscopy (ExM) [10]. It expands samples physically to resolve structures which are unresolvable directly. Besides, there are some other types of super-resolution technique, e.g., fluctuation-based and computer-vision-based approaches [11-13].

By now, super-resolution techniques have not only break the diffraction-limit, but also improved resolutions significantly. In techniques such as STED, PALM, STORM, Confocal, etc., luminous points are distant from one another (or there might be only one luminous point at one time). Their images (almost) do not overlap, thereby the locations and light intensities of these points can be extracted from the blurred image. Inspired by these techniques, we find a "resolvable condition" (please refer to METHODS for its definition) relevant to their imaging conditions. In the "resolvable condition", structures (both inter-points and inner-points) smaller than the diffraction-limit can be extracted directly from the blurred image, even if the points within the structures are imaged at the same time. From the point view of frequency, the image's high frequency part is filtered out by the microscope. But the structures' full information (including both profile and details) could still be recovered from the low frequency part, in the "resolvable condition". Early researches have discussed some mathematic theory for recovering signals from low pass data [14], while this study finds that the detail information is not damaged by diffraction, in the "resolvable condition"; then we proposed a technique to recover a diffraction-blurred image's Region of Interest (ROI) under this condition, in spatial domain and in frequency domain respectively.

## 2. Methods

### 2.1. Background analysis

This study finds a condition and solves a problem in the field of optics. But the methods are partly based on computer science and information technique. Thereby, some background knowledge needs to be introduced briefly. First of all, an appropriate model should be chosen to represent images. In this study, we adopt a classic model widely used in the field of Digital Image Processing [15]. An image is divided into several uniform grids, then each grid is treated as a pixel, and its light intensity is called a pixel value. As a result, the image is represented as a matrix. The matrix (digital image signal) is an approximation of physical image in given sampling rate and quantization accuracy. The structure information of samples, which is what people concern, is carried in the corresponding digital image signals. In this case, our task is not to localize luminous points accurately, but to figure out the light intensity (pixel value) of each grid instead.

The core of this study is based on the following phenomenon: information can be carried in the same signal in different ways. There is a common opinion in the field of Digital Image Processing:

a sample's profile information, which changes slowly in space, corresponds to the low frequency part of its image's Fourier spectrum; while its detail information, which changes fast in space, corresponds to the high frequency part. This is true in usual imaging condition because each pixel value corresponds directly to a grid in image area. Thereby, the spatial structure information is carried directly in pixel values. However, the situation might be different if the information is carried indirectly. Strictly speaking, both high frequency and low frequency components are concepts attached to signals rather than information. They do demonstrate fast-changing or slow-changing forms in space domain. But they might not necessarily correspond to the profile or details of a sample if the information is not carried directly.

Here are some simplified examples about information and its carriers. Example 1: if two physical points are used to carry information, their amount could represent the integer-value "2", or their distance could represent a real-value such as 123.822253. In this example, the information carriers are physical objects. In many other cases, observed signals are used to carry information. Example 2: in a Single-Molecule-Localization microscope, the observed image of individual molecules is blurred, and the pixel values do not show the molecules' detailed structure directly. But what people concern are the molecules' locations and light intensities carried by the pixels indirectly. Such information can be extracted, with methods such as data fitting, when the microscope's Point Spread Function (PSF) is known. In both of the examples, prior knowledge plays a key role, and determines how the information is carried in the signals. In example 1, it tells whether the information is carried in the amount or the distance of the two points. In example 2, it provides the template required for data fitting, i.e., the PSF. Besides the above examples, there are more researches relevant to how information is carried in signals in indirect or implicit ways.

We find that in a certain condition, observed images always carry the full information of a sample's structure, no matter they are sharp or diffraction-blurred. Therefore, the condition is termed "resolvable condition" here, and it has two aspects.

The first aspect is named *isolated lighting* (or *separated lighting*). It means the Region of Interest (ROI) in the sample's image is only affected by its own structure and lighting, and is independent of the rest of the sample and the whole surrounding. For example, only one small area of the sample is lighted, or only one molecule is turned on, while the rest part and the surrounding are either totally dark or have no light collected by the microscope. In practice, an ROI is treated to fulfill isolated lighting as long as the effect of the rest part and the surrounding is ignorable. For example, all the other light sources are far enough away from the ROI, similar to what happens in some super-resolution techniques but may need to be stricter. Such a condition is not difficult to implement with existing techniques. But it actually provides very strong prior knowledge because it determines infinitely many pixel values (i.e., zeros) outside the ROI.

The convolution in a light microscope is usually expressed by the following equation:

$$Observed\ image = Ideal\ image * PSF + noise$$

But the observed part of the sample, i.e., the ROI only has limited size in real applications. Thereby it is also affected by any extra light from other parts or the whole surrounding, especially the structures around the ROI. In this case, the above equation should be modified as follows:

$$Observed\ image = Ideal\ image * PSF + extra + noise$$

Where, "*extra*" means the extra light from other parts or the whole surrounding. In usual case, it is an unknown signal, and its values might be large and affect the result significantly. It tends to be larger when zooming factor is larger and points' images overlap more severely. Actually, the extra light could even be much greater than the true signal itself when zooming factor is very large. For example, **Fig. 1a** and **1b** show samples in normal condition and isolated lighting conditions, respectively. The dashed-line rectangles in all figures indicate the ROI, which is $50 \times 50$ pixels. It simulates a physical region of $100 \times 100$ nm. Thereby, the Airy-disk is about $100 \times 100$ pixels, i.e., $200 \times 200$ nm under this zooming factor. Then, **Fig. 1c** and **1d** is the convolution results of **1a** and **1b**, respectively. The ROI's light intensities in **1c** are overlapped by the images of outside structures. But that in **1d** is affected by nothing else outside the ROI because the surroundings are all dark. As a result, the ROI's light intensities in **1c** are much greater than that in **1d**. Actually, quantitative analysis shows that the former is more than 8.87 times greater than the latter. That means the extra light (unknown) is more than 7.87 times greater than the true signal, and submerges it overwhelmingly. Thereby, the ideal image cannot be figured out from the equation even when there is no noise at all. But in the condition of *isolated lighting*, e.g., in **Fig. 1d**, the extra light is zero. Therefore, a significant barrier on this approach is entirely eliminated.

The second aspect is named *positive effective PSF*, which means that all the values of the effective PSF are positive (i.e., greater than zeros). Where, effective PSF means the part of PSF which affects the convolution results in the ROI. This aspect might be fulfilled in various situations. For example, the effective PSF values are, of course, positive if the PSF is totally positive; this is a little stricter than the situation in usual applications [16]. Or, the PSF may have non-positive values at its "dark rings", but only the central part of the PSF (i.e., the Airy disk) affects the convolution results in the ROI when the ROI is smaller than the diffraction-limit. In this case, the other part of the PSF would only affect the convolution results outside the ROI. Thereby, the effective PSF is the central part, whose values are usually all positive, for normal light microscopes. In practice, approximate solutions might be figured out sometimes even if the conditions are not fulfilled strictly, but the effectiveness would be uncertain.

The pixels of sharp images carry the full information directly, including both profile and details. The blurred images carry not only the profile information directly, but also the full information indirectly in the "resolvable condition". Such a situation of "one carrier, two types of information" is somewhat similar to the above example 1. Different ways of carry lead to different methods for extraction. Full information can be observed directly in sharp images. (Strictly speaking, microscopic images cannot be fully "sharp" because they are always the results of convolution due to diffraction. Fortunately, errors are ignorable when points are distant enough from one another). But more steps might be required to extract full information from blurred images, e.g., solving a system of equations. This study is based on the aforementioned image model, no matter for space domain or frequency domain. Therefore, the task of information extraction is translated into the calculation of unknown pixel values, i.e., matrix elements. The rough locations of unknown pixels in the images should be estimated first, and this could be done using existing techniques such as Single-Molecule-Localization.

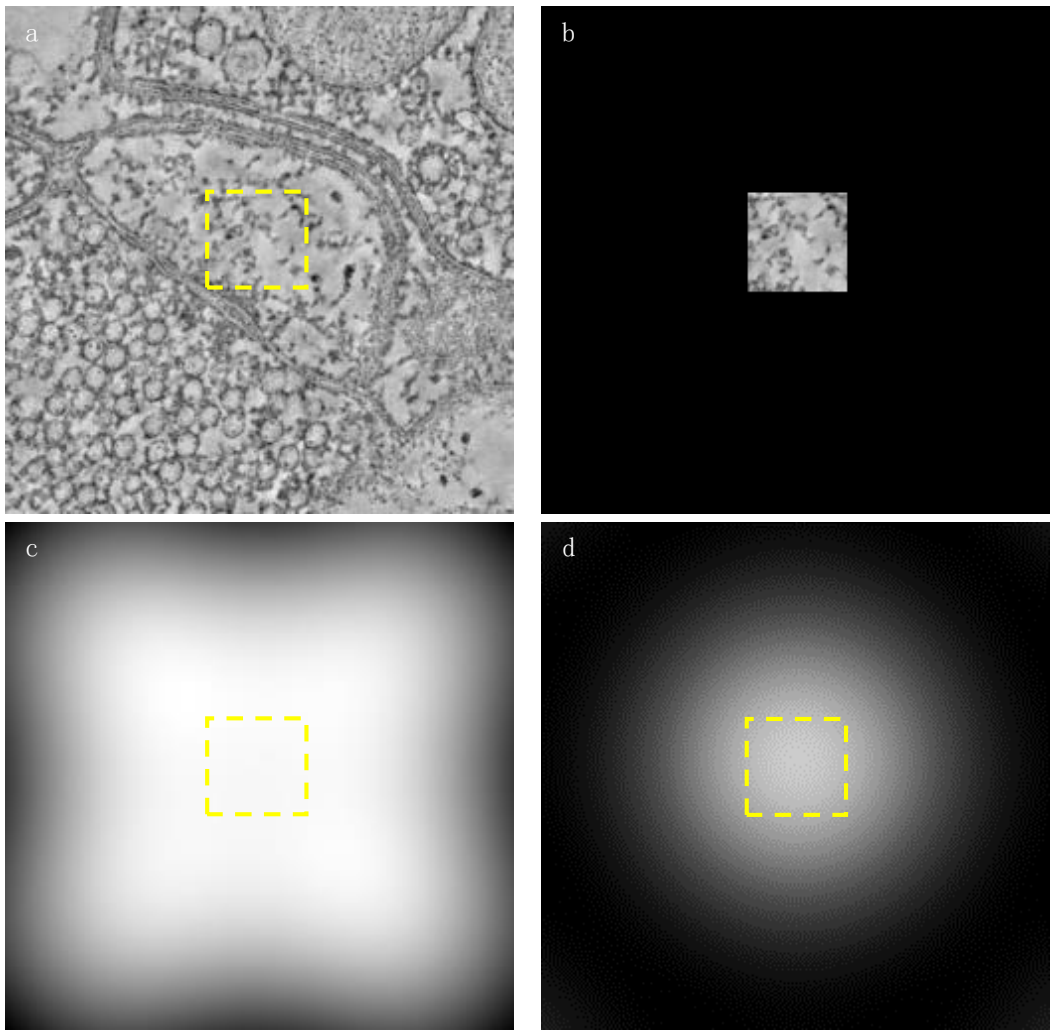

**Fig. 1**. The effect of extra lights on ROI's convolution results. (a) A sample in normal condition. (b) A sample in isolated lighting condition. (c) The convolution result in normal condition. (d) The convolution result in isolated lighting condition. The three dashed-line rectangles indicate the ROIs.

The above phenomenon can also be explained in another way. Assume that a signal (e.g., an image) undertakes a certain processing (e.g., convolution). The ideal (original) signal can be recovered from the result if the processing is reversible. However, a conventional light microscope filters out the high frequency part of any image, thereby the ideal image cannot be recovered reversibly. As mentioned before, the low frequency part of a signal represents its profile, and the high frequency part represents its details. Therefore, details (smaller than the diffraction-limit) are lost after imaged by the microscope. But the processing (convolution caused by the microscope) is reversible in the "resolvable condition". As a result, the details can be recovered from the blurred image or the low frequency part. In the following sections, two recovering methods will be described, for space domain and frequency domain respectively. Noises are neglected here and considered separately.

## 2.2. Method for spatial domain

The effect of diffraction on imaging is usually modeled as the convolution of PSF with ideal images. It diffuses the light intensity of each pixel to other pixels, and thereby lowers the diversity of pixel values, i.e., blurs the images. For convenience, we first explain the method on 1D signals. Analogous to the aforementioned image model, a 1D range is also divided into several uniform segments, and each segment is represented by a value. Let's take a simple signal as an example, as shown by **Fig. 2**. It has two values which are greater than zero, while all the other values are zeros. This is an analogy to the situation that there are only two luminous point.

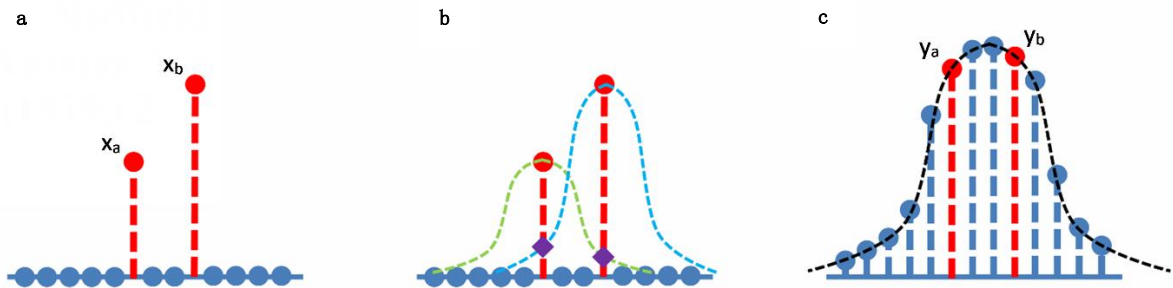

**Fig. 2**. The 1D situation of the spatial domain method. (a) Before convolution (the ideal signal). (b) During convolution. (c) After convolution (the observed signal).

Where, **Fig. 2a** shows the 1D signal before convolution, which named "ideal signal" here. The two values in dash line need to be figured out. **Fig. 2b** shows the situation during convolution, where the convolution kernel is already known. It is equivalent to the PSF of a 2D imaging system, and called Impulse Response Function (IRF) here. **Fig. 2c** shows the resulting signal after convolution. It is also known already, and named "observed signal" here. The result of convolution, i.e., the observed signal looks like the IRF. It is much smoother than the ideal signal, which comprises two impulses. For image signals, being smoother usually means being more blurred, and harder to identify their details.

However, detail information can be recovered from the observed signal, in the "resolvable condition". For the 1D situation in **Fig. 2**, the ROI covers the two non-zero values (just like two separated pixels in an image). The condition means the observed signal's values in the ROI are only relevant to the ideal signal's values in the ROI, and the IRF's values. No other values of the ideal signal can affect the convolution result because they are all zeros. In this case, the observed signal is the convolution of the ideal signal with the IRF. Therefore, the two unknown values in **Fig. 2a** can be figured out from the known IRF and observed signal. (Otherwise, if the "resolvable condition" is not fulfilled, there might be unknown luminous points at unknown locations outside the ROI. Their images, which extend infinitely, would overlap the ROI. Thereby, the aforementioned relationship would be incorrect). Denote:

1. The ideal signal's unknown values are $x_a$ and $x_b$ at the left and right, respectively;
2. The amplitude, i.e., the central value of the IRF is $p$;
3. The IRF has value $q_a$ at the location of $x_a$ when its center is at the location of $x_b$. Therefore, the value $q_a \cdot x_b$ is shown by the left dark diamond in **Fig. 2b**.
4. The IRF has value $q_b$ at the location of $x_b$ when its center is at the location of $x_a$. Therefore, the value $q_b \cdot x_a$ is shown by the right dark diamond in **Fig. 2b**.
5. The observed signal has values $y_a$ and $y_b$ at the location of $x_a$ and $x_b$, respectively.

Since the observed signal is the convolution of the ideal signal with the IRF, we get the following system of equations:

$$\begin{cases} y_a = p \cdot x_a + q_a \cdot x_b \\ y_b = p \cdot x_b + q_b \cdot x_a \end{cases} \quad (1)$$

The solution of the above system of equations is:

$$\begin{cases} x_a = \frac{p \cdot y_a - q_a \cdot y_b}{p^2 - q_a \cdot q_b} \\ x_b = \frac{y_a}{q_a} + \frac{p \cdot q_a \cdot y_b - p^2 \cdot y_a}{p^2 \cdot q_a - q_a^2 \cdot q_b} \end{cases} \quad (2)$$

When $p = 1$ and the IRF is symmetrical, i.e., $q_a = q_b = q$, formula (2) becomes:

$$\begin{cases} x_a = \frac{y_a - q \cdot y_b}{1 - q^2} \\ x_b = \frac{y_b - q \cdot y_a}{1 - q^2} \end{cases} \quad (3)$$

For example, suppose that the ideal signal is $(\cdots \quad 1.2 \quad 3.4 \quad \cdots)$, where the suspension points mean infinitely many zeros; the IRF is $(\cdots \quad 0.9981 \quad 1.0 \quad 0.9981 \quad \cdots)$, where the suspension points mean arbitrary values (they do not affect the result in this case); therefore, $p = 1$, $q_a = q_b = 0.9981$; then assume that the observed signal is $(\cdots \quad 4.5934 \quad 4.5977 \quad \cdots)$, where 4.5934 and 4.5977 are the values of $y_a$ and $y_b$ respectively, and suspension points mean the other values. Substitute these values into formula (2) and we get: $x_a = 1.2$ and $x_b = 3.4$. In other words, the ideal signal is recovered from the observed signal and the IRF.

It can be seen that only part of the observed signal and the IRF is used. In fact, it is not necessary to use $y_a$ and $y_b$ which have the corresponding location to $x_a$ and $x_b$. The method also works if values at other locations are chosen from the observed signal. In practice, it may be helpful for relieving the effect of observation errors if more values are used to build an overdetermined system of equations.

Now we will extend the procedure to 2D signals, e.g., the situation of 2D imaging, as shown by **Fig. 3**. Where, **Fig. 3a** shows the image in ideal conditions, i.e., without the effect of diffraction. Such an image is named "ideal image" in this case. **Fig. 3b** shows the image observed by the microscope, which is named "observed image". It is the convolution result of the ideal image with the PSF, due to diffraction.

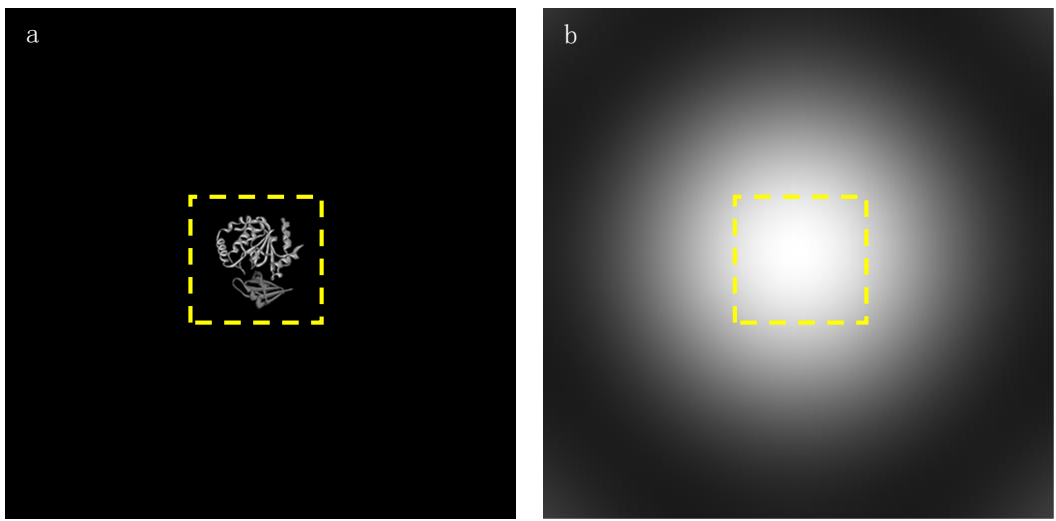

**Fig. 3.** The 2D situation of the spatial domain method. (a) Before convolution (the ideal image). (b) After convolution (the observed image). The two dashed-line rectangles indicate the ROI.

It can be seen that all the pixels in the ideal image equal zeros except in a rectangular ROI. Therefore, the condition of isolated lighting is fulfilled. The PSF's Fourier spectrum is an ideal low pass filter, therefore the PSF extends infinitely in space domain. But the PSF's energy or light intensity is mainly concentrated in the central area. The observed image is extremely blurred, and looks similar to the PSF. It is very difficult to see any details in the observed image, especially in the ROI, which is indicated by a dashed-line rectangle. In fact, the condition of isolated lighting does not restrict the ROI's shape. Even an ROI with many disconnected and irregular areas is acceptable, as long as its rough location could be estimated in the observed image. However, one easy way is to find a rectangle to cover all the areas, and then treat the rectangle as the ROI. In order to decrease the complexity of calculation, the rectangle should be as small as possible.

Similar to 1D situation, the observed image's pixels in the ROI are only relevant to the ideal image's pixels in the ROI and the pixels (values) of PSF. The other pixels in the ideal image do not affect the result of convolution because they are all zeros. In this case, the observed image is the convolution of the ideal image with the PSF. Therefore, the unknown pixels in the ideal image can be figured out from the known PSF and observed image. Let's:

1. Treat the ideal image's ROI as an image named $f(k, l)$, $k = 1, 2, \cdots, K$ and $l = 1, 2, \cdots, L$; where $K$ and $L$ are the amount of the ROI's row and column. $k$ and $l$ start from $(1, 1)$ at the ROI's most left-top pixel.
2. Denote the PSF as image $p(u, v)$, and set a coordinate with its origin at the PSF's center. Thereby, the pixel value is $p(0,0)$ at the PSF's center, and both $u$ and $v$ belong in $[-\infty, +\infty]$;
3. Treat the observed image's ROI as an image named $g(m, n)$, where $m = 1, 2, \cdots, K$ and $n = 1, 2, \cdots, L$.

Let's take a rotationally symmetrical PSF as an example, and the other PSF could be handled similarly. During convolution, let the PSF overlap the ideal image, and align the PSF's center with each pixel in the ideal image's ROI each time. Then, multiply each pixel in the ideal image's ROI with its corresponding PSF pixel, and accumulate all the results. The accumulative value is the observed image's pixel value in the corresponding location. The above procedure could be implemented concisely with a program. In mathematics, this could be expressed with a system of linear equations as follows:

$$Ax = y \tag{4}$$

Where:

$$A = \begin{pmatrix} p(0,0) & \cdots & p(0, L-1) & \cdots & p(K-1, 0) & \cdots & p(K-1, L-1) \\ \vdots & \ddots & \vdots & \ddots & \vdots & \ddots & \vdots \\ p(0, -L+1) & \cdots & p(0,0) & \cdots & p(K-1, -L+1) & \cdots & p(K-1, 0) \\ \vdots & \ddots & \vdots & \ddots & \vdots & \ddots & \vdots \\ p(-K+1, 0) & \cdots & p(-K+1, L-1) & \cdots & p(0,0) & \cdots & p(0, L-1) \\ \vdots & \ddots & \vdots & \ddots & \vdots & \ddots & \vdots \\ p(-K+1, -L+1) & \cdots & p(-K+1, 0) & \cdots & p(0, -L+1) & \cdots & p(0, -0) \end{pmatrix}$$

This is a matrix with a size of $(K \cdot L) \times (K \cdot L)$. Then:

$$x = \begin{pmatrix} f(1,1) \\ \vdots \\ f(1, L) \\ \vdots \\ f(K, 1) \\ \vdots \\ f(K, L) \end{pmatrix}$$

This is a matrix (vector) with a size of $(K \cdot L) \times 1$, and it is actually a sequence of all the pixels in the ideal image's ROI, arranged row by row from top to bottom. Then:

$$y = \begin{pmatrix} g(1,1) \\ \vdots \\ g(1,L) \\ \vdots \\ g(K,1) \\ \vdots \\ g(K,L) \end{pmatrix}$$

This is also a matrix (vector) with a size of $(K \cdot L) \times 1$, and it is actually a sequence of all the pixels in the observed image's ROI, arranged row by row from top to bottom.

The above $A$ is determined by the PSF, and $y$ is determined by the observed image. In other words, both $A$ and $y$ are already known, and $x$ is actually the rearrangement of the ideal image's unknown pixels. In theory, if we substitute the ideal image's vector $x$ into formula (4), the resulting $Ax$ should equals $y$. Accordingly, the ideal image can be get by solving formula (4). In the above procedure, only the pixels in the observed image's ROI are adopted. Actually, if the other pixels of the observed image are also used, an overdetermined system could be build including more equations. In practice, that may be helpful for improving the method's capability of noise resistance. In this case, the unknowns are still the ideal image's pixels in the ROI because all the other pixels are known to be zeros, in the condition of isolated lighting.

There are many classic or cutting-edge methods can be adopted for solving the above system of equations. Its solvability can be explain as follows. On the one hand, the system of equations should obviously have at least one solution, i.e., the ideal image itself. On the other hand, the "resolvable condition" has a second aspect, i.e., effective PSF is totally positive. Therefore, the elements of matrix $A$ are all positive. In addition, the elements of vector $x$, i.e., the ideal image's pixels are all non-negative also, thereby $Ax = 0$ is true only when $x = 0$. In other words, the system of homogeneous linear equations $Ax = 0$ only has the zero solution. Therefore, according to the property of system of linear equations, the corresponding system of nonhomogeneous linear equations $Ax = y$ has only one solution (i.e., the ideal image) [17].

In the above procedure, the ideal image is figured out by building and solving a system of equations. This seems to be unreasonable in frequency domain because the high frequency part of the image has been filtered out by the microscope. However, we find that the full information of a sample's spatial structure can be recovered even with only a small amount of frequency elements in the "resolvable condition".

### 2.3. Method for frequency domain

According to Fourier Optics, convolution caused by diffraction is equivalent to filtering ideal images with an ideal low pass filter. Assume that the low pass filter's amplitude is 1, without loss of generality. First, let's describe the method on a simple 1D signal, as shown by **Fig. 4**.

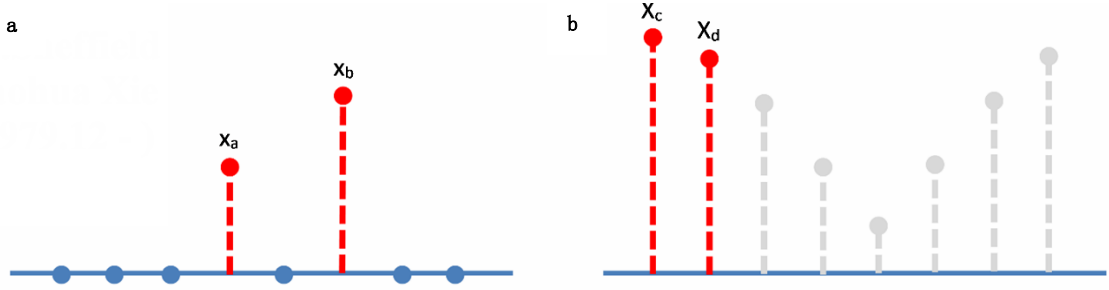

**Fig. 4.** The 1D situation of the frequency domain method. (a) The ideal signal including only two non-zero values. (b) The corresponding spectrum, where only the most left two values are preserved.

Where, **Fig. 4a** shows an ideal signal. It has only two unknown values, while all the other values are zeros. Therefore, the condition of isolated lighting is fulfilled. **Fig. 4b** shows the ideal signal's Fourier spectrum. After low pass filtering, all the high frequency components become zeros. Therefore, the known spectrum only includes low frequency components, and is named "observed spectrum" here. For example, the most left two values in the spectrum are preserved, and treated as the observed spectrum. In 1D case, the low pass filter is equivalent to the IRF's Fourier spectrum, which is usually called System Transfer Function (STF). In this case, there is a mathematical relationship between the ideal signal and the observed spectrum. Assume that:

1. The length of the ideal signal is $N$, and $N \geq 2$;
2. The ideal signal's unknown values are $x_a$ and $x_b$ respectively, where both $a$ and $b$ are integers within $[0, N-1]$;
3. Arbitrarily choose two values $X_c$ and $X_d$ from the observed spectrum, where both $c$ and $d$ are integers.

The formula of 1D discrete Fourier Transform is as follows [18]:

$$X_k = \sum_{n=0}^{N-1} x_n \cdot e^{-\frac{2\pi i}{N} \cdot k \cdot n} \tag{5}$$

Since all the ideal signal's values are zeros except $x_a$ and $x_b$, the above formula becomes:

$$X_k = x_a \cdot e^{-\frac{2\pi a k i}{N}} + x_b \cdot e^{-\frac{2\pi b k i}{N}} \tag{6}$$

Let $k$ equals $c$ and $d$ respectively, then substitute them into formula (6), and we get:

$$\begin{cases} X_c = x_a \cdot e^{-\frac{2\pi a c i}{N}} + x_b \cdot e^{-\frac{2\pi b c i}{N}} \\ X_d = x_a \cdot e^{-\frac{2\pi a d i}{N}} + x_b \cdot e^{-\frac{2\pi b d i}{N}} \end{cases} \tag{7}$$

Solve the above system of linear equations, and we get:

$$x_a = \frac{X_c \cdot e^{-\frac{2\pi b d i}{N}} - X_d \cdot e^{-\frac{2\pi b c i}{N}}}{e^{-\frac{2\pi(ac+bd)i}{N}} - e^{-\frac{2\pi(ad+bc)i}{N}}} \tag{8}$$

and:

$$x_b = \frac{X_c - x_a \cdot e^{\frac{-2\pi a c i}{N}}}{e^{\frac{-2\pi b c i}{N}}} \tag{9}$$

For example, assume that the ideal signal is $(0 \quad 0 \quad 0 \quad 6.7 \quad 8.9 \quad 0 \quad 0 \quad 0)$, and thereby the corresponding Fourier spectrum is $(15.6 \quad -13.6376 - 4.7376i \quad \cdots)$. In the latter, the complex values $15.6$ and $-13.6376 - 4.7376i$ are the chosen components of the observed spectrum, and the suspension points represent the other components. Therefore, $N = 8$, $a = 3$, $b = 4$, $c = 0$, $d = 1$, $X_c = 15.6$, $X_d = -13.6376 - 4.7376i$. Please note that $a, b, c, d$ are all coordinate indices of 1D signals, which start from 0. Substitute the above values into formulas (8) and (9), and we get $x_a = 6.7$ and $x_b = 8.9$. In other words, the ideal signal is recovered from the observed (low frequency) spectrum.

It can be seen that only two frequency components of the observed spectrum are used. Actually, they could be chosen from the low frequency part, or the high frequency part, or even arbitrary part. This method works as long as a system of equations is built and a unique solution is got. In practice, it may be helpful for relieving the effect of errors if more frequency components are used to build an overdetermined system of equations.

Now we will extend the procedure to 2D signals, e.g., the situation of 2D imaging. In this case, the STF is called Optical Transfer Function (OTF) which is equivalent to the PSF's Fourier spectrum, as shown by **Fig. 5**.

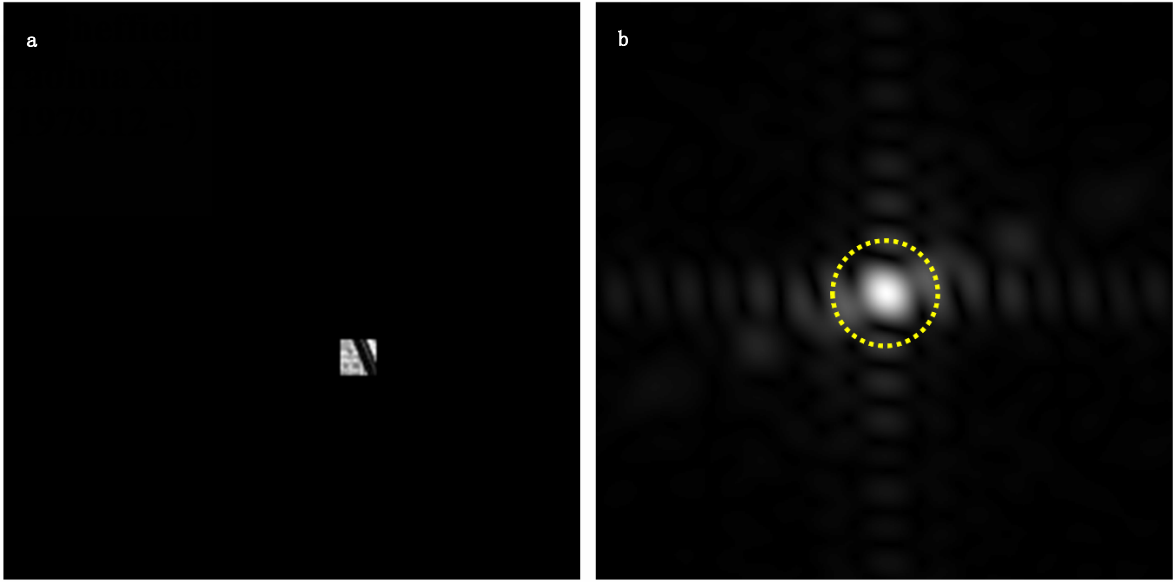

**Fig. 5.** The 2D situation of the frequency domain method. (a) The ideal image including a small ROI. (b) The corresponding frequency spectrum, where the low frequency part is encircled.

Where, **Fig. 5a** shows an ideal 2D signal without being filtered, which is named "ideal image" here. **Fig. 5b** shows the ideal image's spectrum of 2D discrete Fourier transform. After being filtered by an ideal low pass filter, the spectrum's components outside the dotted circle are all removed. Thereby, the known spectrum includes components inside the circle only, and is named "observed spectrum". Please note that the spectrum is shown conventionally in **Fig. 5b**, i.e., the low frequency components are at the center. But it is still handled in original format in the following procedure, i.e., the high frequency components are at the center. In this case, there is a mathematical relationship between the ideal image and the observed spectrum. Assume that:

1. The ideal image is $x(m,n)$; $m = 1, 2, \cdots, M$, $n = 1, 2, \cdots, N$, where $M$ and $N$ are the amount of the image's row and column, respectively; $m$ and $n$ start from $(1,1)$ at the image's most left-top pixel;
2. The size of the ideal image's ROI is $K \times L$, and the row number and the column number of the ROI's left-top pixel are $a$ and $b$, respectively;
3. The ideal image's full spectrum is $Y(u,v)$, where $u = 1, 2, \cdots, M$, $v = 1, 2, \cdots, N$;
4. Choose a rectangular area from the observed spectrum, with a size of $K \times L$; the row number and the column number of its left-top pixel are $c$ and $d$, respectively; for example, $c = d = 0$.

The formula of 2D discrete Fourier transform is as follows [15]:

$$Y(u,v) = \frac{1}{MN} \cdot \sum_{m=0}^{M-1} \sum_{n=0}^{N-1} x(m,n) \cdot e^{-2\pi i \left(\frac{m \cdot u}{M} + \frac{n \cdot v}{N}\right)} \tag{10}$$

Since all the ideal image's pixels are zeros except those within the ROI, the above formula becomes:

$$Y(u,v) = \frac{1}{MN} \cdot \sum_{m=a}^{a+K-1} \sum_{n=b}^{b+L-1} x(m,n) \cdot e^{-2\pi i \left(\frac{m \cdot u}{M} + \frac{n \cdot v}{N}\right)} \tag{11}$$

Substitute the chosen components of the observed spectrum into formula (11), and we get a system of equations $Ax = y$, where:

$$A = \frac{1}{MN} \begin{pmatrix} e^{-2\pi i \left(\frac{a \cdot c}{M} + \frac{b \cdot d}{N}\right)} & \cdots & e^{-2\pi i \left(\frac{a \cdot c}{M} + \frac{(b+L-1) \cdot d}{N}\right)} & \cdots & e^{-2\pi i \left(\frac{(a+K-1) \cdot c}{M} + \frac{b \cdot d}{N}\right)} & \cdots & e^{-2\pi i \left(\frac{(a+K-1) \cdot c}{M} + \frac{(b+L-1) \cdot d}{N}\right)} \\ \vdots & \ddots & \vdots & \ddots & \vdots & \ddots & \vdots \\ e^{-2\pi i \left(\frac{a \cdot c}{M} + \frac{b \cdot (d+L-1)}{N}\right)} & \cdots & e^{-2\pi i \left(\frac{a \cdot c}{M} + \frac{(b+L-1) \cdot (d+L-1)}{N}\right)} & \cdots & e^{-2\pi i \left(\frac{(a+K-1) \cdot c}{M} + \frac{b \cdot (d+L-1)}{N}\right)} & \cdots & e^{-2\pi i \left(\frac{(a+K-1) \cdot c}{M} + \frac{(b+L-1) \cdot (d+L-1)}{N}\right)} \\ \vdots & \ddots & \vdots & \ddots & \vdots & \ddots & \vdots \\ e^{-2\pi i \left(\frac{a \cdot (c+K-1)}{M} + \frac{b \cdot d}{N}\right)} & \cdots & e^{-2\pi i \left(\frac{a \cdot (c+K-1)}{M} + \frac{(b+L-1) \cdot d}{N}\right)} & \cdots & e^{-2\pi i \left(\frac{(a+K-1) \cdot (c+K-1)}{M} + \frac{b \cdot d}{N}\right)} & \cdots & e^{-2\pi i \left(\frac{(a+K-1) \cdot (c+K-1)}{M} + \frac{(b+L-1) \cdot d}{N}\right)} \\ \vdots & \ddots & \vdots & \ddots & \vdots & \ddots & \vdots \\ e^{-2\pi i \left(\frac{a \cdot (c+K-1)}{M} + \frac{b \cdot (d+L-1)}{N}\right)} & \cdots & e^{-2\pi i \left(\frac{a \cdot (c+K-1)}{M} + \frac{(b+L-1) \cdot (d+L-1)}{N}\right)} & \cdots & e^{-2\pi i \left(\frac{(a+K-1) \cdot (c+K-1)}{M} + \frac{b \cdot (d+L-1)}{N}\right)} & \cdots & e^{-2\pi i \left(\frac{(a+K-1) \cdot (c+K-1)}{M} + \frac{(b+L-1) \cdot (d+L-1)}{N}\right)} \end{pmatrix}$$

This is a matrix with a size of $(K \cdot L) \times (K \cdot L)$. Then:

$$x = \begin{pmatrix} x(a,b) \\ \vdots \\ x(a, b+L-1) \\ \vdots \\ x(a+K-1, b) \\ \vdots \\ x(a+K-1, b+L-1) \end{pmatrix}$$

This is a matrix (vector) with a size of $(K \cdot L) \times 1$, and it is a sequence of all the pixels in the ideal image's ROI, arranged row by row from top to bottom. Then:

$$y = \begin{pmatrix} y(c,d) \\ \vdots \\ y(c,d+L-1) \\ \vdots \\ y(c+K-1,d) \\ \vdots \\ y(c+K-1,d+L-1) \end{pmatrix}$$

This is also a matrix (vector) with a size of $(K \cdot L) \times 1$, and it is a sequence of the chosen components of the observed spectrum, arranged row by row from top to bottom.

Similar to the method for spatial domain, vector $x$ can be got by solving $Ax = y$, and then the ideal image can be got by the rearrangement of $x$. Although the rectangular area are chosen from the observed spectrum, other shapes or even randomly chosen components are also allowed in this method. This method works as long as a system of equations is built and a unique solution is got. In practice, it may be helpful for relieving the effect of errors if more frequency components are used to build an overdetermined system of equations.

It can be seen that larger ROI could be recovered if more spectrum components are preserved after filtering. In a special (extreme) case, all the spectrum components are known. Thereby, the ideal image could be got directly by inverse filtering or inverse Fourier transform. Actually, it is a classic way recovering a full image from its full Fourier spectrum. This study finds more possibility: recovering part of an image (with full details) from part of its Fourier spectrum (e.g., including only low frequency). That means even low frequency components carry full details of a sample's spatial structure (in an ROI), and seems inconsistent with traditional opinions. After excluding several explanations, we believe that the reason is the "integrity of spectrum", i.e., different frequency components are tightly relevant in the "resolvable condition". Let's take 2D case as an example. As can be seen from the formula of 2D discrete Fourier transform, the spectrum is actually the accumulation of the products of each pixel $x(m,n)$ with its corresponding basis function in frequency domain. Each product is as follows:

$$\frac{1}{MN} \cdot x(m,n) \cdot e^{-2\pi i \left(\frac{m \cdot u}{M} + \frac{n \cdot v}{N}\right)}$$

This is a function including all frequency components, and its amplitude is affected by the corresponding pixel value $x(m,n)$. When the pixel value varies, the function's values change accordingly at any frequency with the same percentage. In other words, each pixel value is carried on the amplitude of its corresponding function, or the function's value carries its corresponding pixel value at each frequency (from the lowest to the highest frequency). Taking an extreme situation as an example, there is only one function (product) in the image's spectrum when there is only one pixel in the image's ROI. In other words, the spectrum is the product of the only pixel with its corresponding basis function. When $M$, $N$ and the pixel's location $(m,n)$ is known, the basis function $\frac{1}{MN} \cdot e^{-2\pi i\left(\frac{m \cdot u}{M}+\frac{n \cdot v}{N}\right)}$ is also known. Therefore, pick the observed spectrum's value at an arbitrarily selected frequency, and divided it by the basis function's value at the same frequency, then the result is the unknown pixel value $x(m, n)$. Actually, the selected frequency could even be zero frequency, i.e., DC component in this case. This situation is similar to that when an individual molecule's light intensity is extracted using Single-Molecule-Localization techniques. When there are more unknown pixels, they can be figured out by building and solving a system of equations, as shown by the aforementioned procedure. From this point of view, this technique could be treated as the extension of existing techniques such as Single-Molecule-Localization, and it further "split" a single point into $2 \times 2$, $3 \times 3$ or more points, i.e., the ROI. It can be seen that the full frequency spectrum is relatively redundant when the ROI is smaller than the ideal image. In this case, there is no need to recover the full spectrum, as what popular deconvolution techniques do, for the recovery of the ideal image. A blurred image without high frequency components looks meaningless and less informative, but it actually contains the full information of the ideal image, in the "resolvable condition". In other words, after different ideal images are filtered, the resulting blurred images all seem similar and undistinguishable. But they are actually different from one another, as can be seen from formula (10). From the information theory's point of view, that means they actually carry different information, i.e., they are distinguishable [19].

## 3. Results and discussion

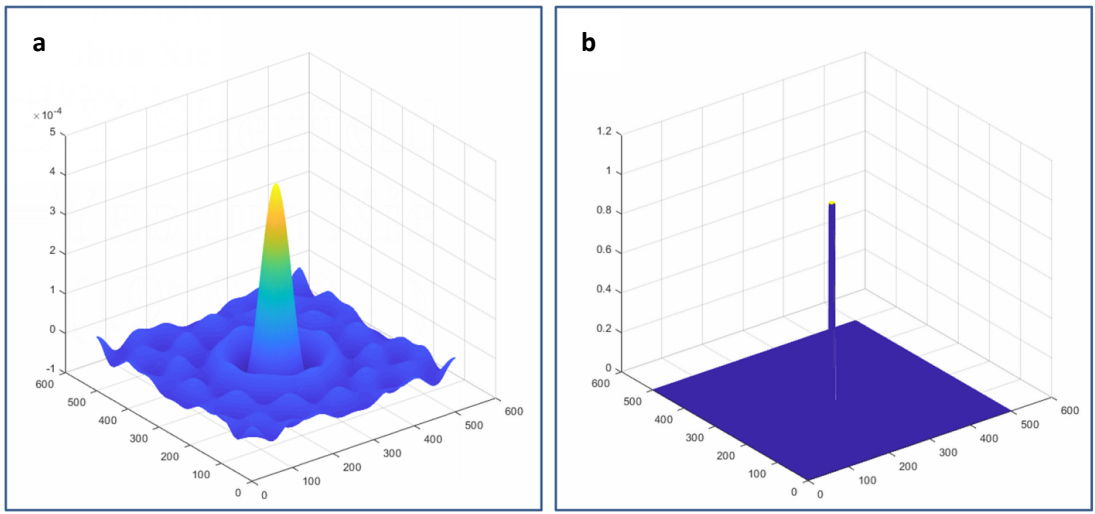

**Fig. 6.** The PSF and its Fourier spectrum. (a) The experimental PSF. (b) The PSF's Fourier spectrum which is an ideal low pass filter. Frequency components are all removed if they are more than 6 pixels away from the center, i.e., zero frequency.

Simulation experiments are performed for spatial domain and frequency domain respectively. Each ideal image (expected image) tested is of size $768 \times 768$, and has an ROI in it. Then, 19 different sizes of ROI, i.e., $2 \times 2$, $3 \times 3$, $\cdots$, $20 \times 20$ pixels, are tested for both spatial domain and frequency domain respectively. Furthermore, 20 random tests are performed for each size. In order to cover various possibilities, the pixel values in the ROI are randomly generated in each test. Therefore, $19 \times 20 = 380$ different ideal images are tested for spatial domain and frequency domain, respectively.

In theory, the Airy-disk-shaped PSF extends infinitely because the microscope is an ideal low pass filter. But according to the analysis in METHODS, the convolution result in the ROI is affected only by the PSF's central area when the PSF is large enough. Therefore, the PSF of size $501 \times 501$ is adopted in this experiment as shown by **Fig. 6a**. The PSF's Fourier spectrum is shown by **Fig. 6b**.

### 3.1. Results for spatial domain

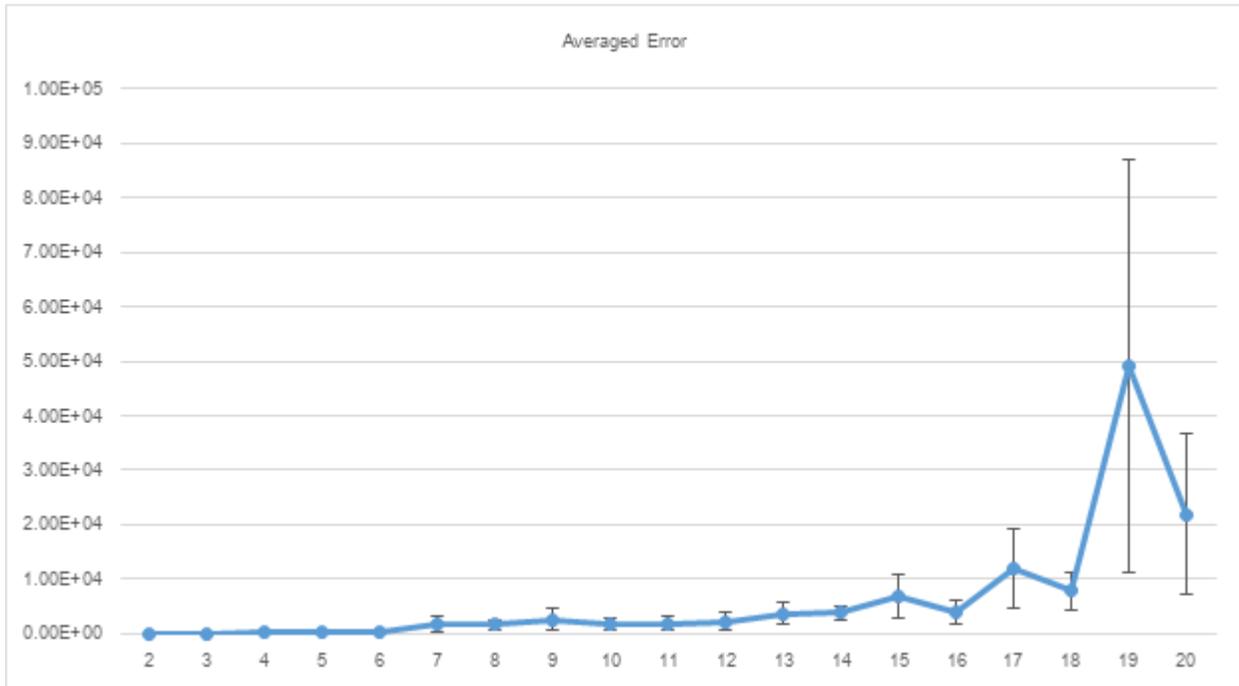

**Fig. 7.** The Averaged Errors of the spatial domain method. Where, lateral axis shows the ROI's size, from $2 \times 2$ to $20 \times 20$ pixels. Vertical axis shows the Averaged Error. The error bars show the standard deviations of the Average Errors. The value 1.00E+05 means $1.00 \times 10^5$, and the other values are similar.

Now we will describe the experiment for spatial domain. After convolution, all the observed images look blurred, just like **Fig. 1d**, but they actually carry information different from one another. Figure out the ideal images' unknown pixels using our method for spatial domain, and the Averaged Errors (AEs) of the results are shown by **Fig. 7**. There is an Averaged Error for each size, which means the average of all the testing errors in the 20 tests. The testing error is defined as $\|x - x'\|/(K \cdot L)$, which represents the mean square error between the recovered pixels and their corresponding pixels in the ideal image's ROI. In the formula, $x$ represents the vector rearranged from recovered pixels, $x'$ represents the vector rearranged from the pixels in the ideal image's ROI, and $K \cdot L$ is the count of pixels in the ROI. All the Averaged Errors are also shown in **Table 1**, as follows.

| ROI size |  | $2 \times 2$ | $3 \times 3$ | $4 \times 4$ | $5 \times 5$ |
|---|---|---|---|---|---|
| AE |  | 3.23E-08 | 0.21 | 202.93 | 236.88 |
| ROI size | $6 \times 6$ | $7 \times 7$ | $8 \times 8$ | $9 \times 9$ | $10 \times 10$ |
| AE | 458.63 | 1778.16 | 1634.61 | 2649.79 | 1652.73 |
| ROI size | $11 \times 11$ | $12 \times 12$ | $13 \times 13$ | $14 \times 14$ | $15 \times 15$ |
| AE | 1909.78 | 2252.56 | 3667.90 | 3863.32 | 6898.65 |
| ROI size | $16 \times 16$ | $17 \times 17$ | $18 \times 18$ | $19 \times 19$ | $20 \times 20$ |
| AE | 3835.98 | 12051.00 | 7861.40 | 49128.85 | 21946.59 |

**Table 1.** The Averaged Errors (AEs) of the spatial domain method. The value 3.23E-08 means $3.23 \times 10^{-8}$, and the others are similar.

It can be seen from **Table 1** that the averaged errors are very tiny for sizes $2 \times 2$ and $3 \times 3$. Given the upper limit of pixel value is 256, the recovered results can be treated as the same as the corresponding ideal images. That verifies the effectiveness of the method: the spatial resolution is increased by 3 times in each dimension if each individual pixel is resolved into $3 \times 3$ pixels with full details. For sizes $4 \times 4$ to $20 \times 20$, the averaged errors are larger, and it is hard to judge whether the method still works in these cases.

Therefore, we calculate the Averaged Difference (ADs). There is an Averaged Difference for each size, which means the average of all the difference values in the 20 tests. Where, the difference value is defined as $\|Ax - y\|/(K \cdot L)$, which reflects how well the vector $x$ fulfills the formula $Ax = y$. In this experiment, $y$ represents the vector rearranged from observed pixels in the ROI. By checking the difference value, we can see how well the ideal image fulfill the corresponding system of equations. For each size of ROI, the difference values are averaged for all the 20 random tests. The resulting Averaged Differences are shown in **Table 2**. In addition, the standard deviations of difference values are less than $1E - 16$ for all the above sizes. According to these results, the system of equations in this method can model the imaging procedure accurately enough. That suggests that the large errors in **Table 1** are not caused by the method's principle. More accurate results could be got if more effective approaches are used to solve the system of equations. Therefore, the method's effectiveness is also proved indirectly for size $4 \times 4$ to $20 \times 20$.

| ROI size |          | $2 \times 2$   | $3 \times 3$   | $4 \times 4$   | $5 \times 5$   |
|----------|----------|----------|----------|----------|----------|
| AD       |          | 2.43E-17 | 2.48E-17 | 3.98E-17 | 4.86E-17 |
| ROI size | $6 \times 6$   | $7 \times 7$   | $8 \times 8$   | $9 \times 9$   | $10 \times 10$ |
| AD       | 7.30E-17 | 1.03E-16 | 1.09E-16 | 1.60E-16 | 2.00E-16 |
| ROI size | $11 \times 11$ | $12 \times 12$ | $13 \times 13$ | $14 \times 14$ | $15 \times 15$ |
| AD       | 2.12E-16 | 2.44E-16 | 3.42E-16 | 4.00E-16 | 4.12E-16 |
| ROI size | $16 \times 16$ | $17 \times 17$ | $18 \times 18$ | $19 \times 19$ | $20 \times 20$ |
| AD       | 4.26E-16 | 4.58E-16 | 5.91E-16 | 6.99E-16 | 7.62E-16 |

**Table 2.** The Averaged Differences (ADs) of the spatial domain method.

### 3.2. Results for frequency domain

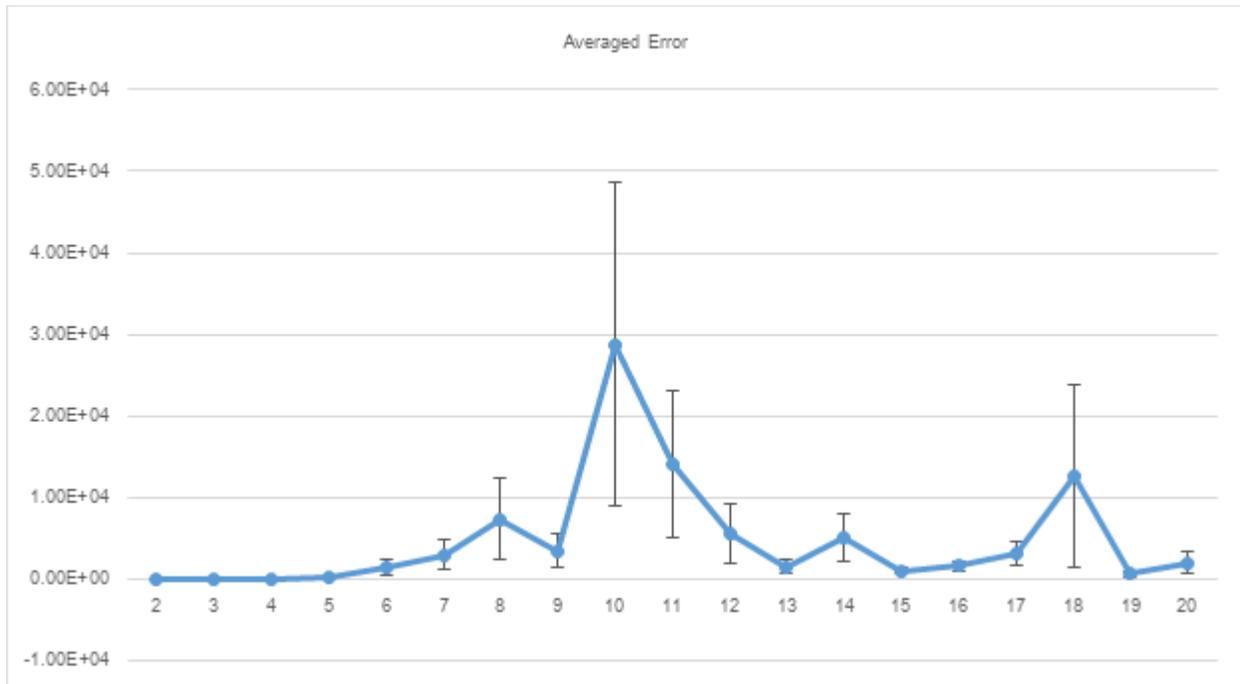

**Fig. 8.** The Averaged Errors of the frequency domain method. Similar to the experiment for spatial domain, the lateral axis shows the ROI's size, and the vertical axis shows the Averaged Errors. The error bars also show the standard deviations of the Average Errors.

Then we will describe the experiment for frequency domain. In this experiment, the observed image's frequency spectrum is zero except its low frequency part. Figure out the ideal image's unknown pixels using our method for frequency domain. The resulting Averaged Errors (AEs) are shown in **Fig. 8**, and **Table 3**.

| ROI size |          | $2 \times 2$ | $3 \times 3$ | $4 \times 4$ | $5 \times 5$ |
|----------|----------|--------------|--------------|--------------|--------------|
| AE       |          | 1.21E-09     | 1.63E-04     | 0.59         | 118.56       |
| ROI size | $6 \times 6$ | $7 \times 7$ | $8 \times 8$ | $9 \times 9$ | $10 \times 10$ |
| AE       | 1432.21  | 2977.23      | 7357.42      | 3455.11      | 28796.84     |
| ROI size | $11 \times 11$ | $12 \times 12$ | $13 \times 13$ | $14 \times 14$ | $15 \times 15$ |
| AE       | 14055.47 | 5611.23      | 1518.55      | 5025.37      | 1011.36      |
| ROI size | $16 \times 16$ | $17 \times 17$ | $18 \times 18$ | $19 \times 19$ | $20 \times 20$ |
| AE       | 1632.88  | 3181.26      | 12717.13     | 641.46       | 2037.52      |

**Table 3.** The Averaged Errors (AEs) of the frequency domain method.

As can be seen from **Table 3**, the Averaged Errors are very tiny for sizes $2 \times 2$ to $4 \times 4$. Therefore, the recovered results can be treated as the same as the corresponding ideal images. That verifies the effectiveness of the method: the spatial resolution is increased by 4 times in each dimension if each pixel is resolved into $4 \times 4$ pixels with full details. The errors for size $4 \times 4$ in the spatial method are larger than those here. But both the two methods are based on the same principle. This suggests that the larger errors are not caused by the principle.

| ROI size |          | $2 \times 2$ | $3 \times 3$ | $4 \times 4$ | $5 \times 5$ |
|----------|----------|--------------|--------------|--------------|--------------|
| AD       |          | 5.86E-14     | 1.72E-13     | 2.23E-13     | 2.86E-13     |
| ROI size | $6 \times 6$ | $7 \times 7$ | $8 \times 8$ | $9 \times 9$ | $10 \times 10$ |
| AD       | 5.11E-13 | 5.63E-13     | 9.39E-13     | 1.13E-12     | 1.39E-12     |
| ROI size | $11 \times 11$ | $12 \times 12$ | $13 \times 13$ | $14 \times 14$ | $15 \times 15$ |
| AD       | 1.58E-12 | 2.20E-12     | 2.51E-12     | 2.79E-12     | 3.18E-12     |
| ROI size | $16 \times 16$ | $17 \times 17$ | $18 \times 18$ | $19 \times 19$ | $20 \times 20$ |
| AD       | 3.46E-12 | 3.77E-12     | 4.03E-12     | 4.15E-12     | 4.37E-12     |

**Table 4.** The Averaged Differences (ADs) of the frequency domain method.

Then, the Averaged Differences (ADs) are also calculated, as shown by **Table 4**. In addition, the standard deviations of difference values are less than $1E - 12$ for all the sizes. According to these results, the system of equations in this method can model the imaging procedure accurately enough. More accurate results could be got if more effective approaches are used to solve the system of equations. Therefore, the method's effectiveness is also proved indirectly for sizes $5 \times 5$ to $20 \times 20$.

Furthermore, the methods are also verified indirectly for size $21 \times 21$ to $100 \times 100$. Given the limitation of computational resource and time, only one random test is performed for each size. All the results demonstrate that the methods can model the imaging procedure precisely because the Averaged Difference is smaller than $5E - 12$ in each test. In practice, the $100 \times 100$ ROI can be a single luminous point such as a fluorescent molecule.

Fig. 9 shows the zoomed view of an experiment result. The $3 \times 3$ pixels at the center of **Fig. 9a** or **Fig. 9c** form the ROI, which is shown as a square at the center of **Fig. 9b**. After convolution, the observed image actually extends infinitely, and it is very blurred. Thereby, the pixels in **Fig. 9b** look almost all the same, but they are different slightly. Existing techniques could be used to estimate the ROI's location, which is indicated by the asterisk in **Fig. 9b**. The recovered image is almost the same as the ideal image, and includes both profile and detail information.

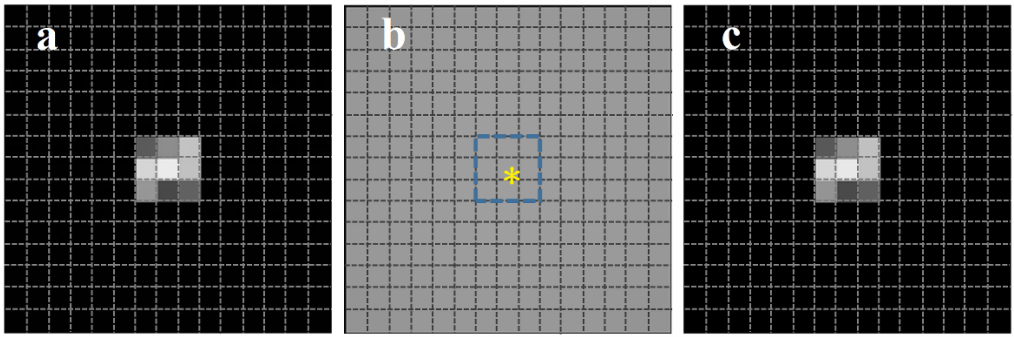

**Fig. 9.** The zoomed view of an experiment result. (a) Ideal image (before convolution). (b) Observed image (after convolution). (c) Recovered image. In all the three images, dot-line grids represent the borders of pixels. The asterisk in (b) indicates the estimated location of the ROI.

An experiment is also performed on larger images, and the result of an example is shown by **Fig. 10**. Where, **Fig. 10a** is the ideal image, which simulates the sample's physical structure including various objects. Although it is a synthesized image ($300 \times 300$ pixels), it is enough to test the proposed technique in principle. **Fig. 10b** is the direct convolution of the ideal image with the PSF, and it simulates the image from a conventional microscope. **Fig. 10c** is the result recovered with the proposed technique. For this simulated microscope, assume that its resolution is 5 nm per pixel. The sample (ideal image) is scanned with a luminous point of $15\text{nm} \times 15\text{nm}$, just like what happens in STED. Thereby, the ROI is $3 \times 3$ pixels at each time, and the ideal image ($300 \times 300$ pixels) is scanned for totally $100 \times 100$ times. The Airy-disk' radius is approximately between 200 to 300 nm, thereby it is between 40 and 60 pixels. The conventional result (**Fig. 10b**) is very blurred (even hard to see meaningful profile) because the Airy-disk is much larger than any details. But the recovered image is sharp, with full details, and the averaged error is ignorable (about 0.0045 of the averaged pixel value of the ideal image). As the resolution of a conventional microscope is about 200nm, the improvement of resolution is about 200/5 = 40 times in this experiment. In principle, the whole image could be treated as an ROI and recovered in one pass, as long as the "resolvable condition" is fulfilled. But in practice, it would be difficult to solve the large system of equations (300 × 300 unknowns in this example) accurately enough.

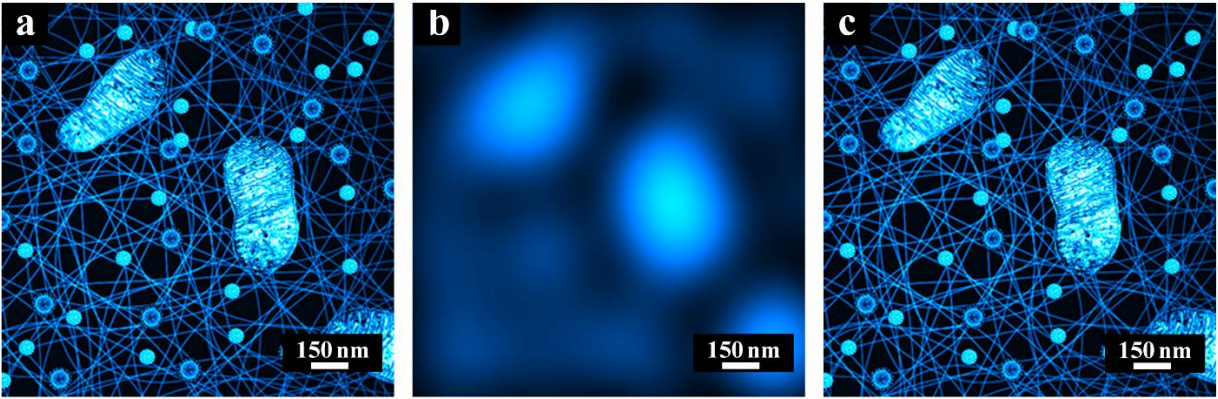

**Fig. 10.** An example of larger images. (a) Ideal image (sample structures). (b) Convolved directly (conventional microscope). (c) Recovered with the proposed technique.

The recovered image would have similar accuracy to the simulation experiment if the observed image and PSF are completely accurate. But noises are almost inevitable in practice. A preliminary experiment shows that the recovered image is acceptable when PSNR (Peak Signal-to-Noise Ratio) is greater than a certain threshold such as 250. It can be seen that the technique is sensitive to noise, thereby further research on SNR improvement would be necessary. There are many valuable methods in fields such as Noise Reduction which may be utilized. Besides, the proposed technique recovers detail information directly from an image without high frequency components. But noises usually contain more high frequency components. This difference may also be helpful to suppress noises.

## 4. Conclusion

Usually, an image's low frequency part represents its profile, and high frequency part represents its details. A conventional light microscope filters out high frequency part, and thereby makes the details cannot be recovered. In other words, the convolution caused by the microscope is irreversible. But this study finds a "resolvable condition" which is relevant to the imaging condition of existing techniques. In this condition, the observed image (or its Fourier spectrum) has redundancy. Thereby, the details can be recovered from part of the observed image, or part of its Fourier spectrum (e.g., low frequency components). In other words, a sample's structure (including both profile and details) can be extracted directly from a blurred image, even if its size is much smaller than the diffraction-limit and its inner points are imaged at the same time. Thereby, the convolution is reversible and the recovery is independent of high frequency in this condition.

Then, a technique is proposed based on the above findings and image deconvolution. The latter is usually employed for the recovery of degraded images, e.g., relieving the effect of defocused light. Usually, it cannot recover the results of ideal low pass filters. But some classic deconvolution approaches are modified to a super-resolution technique when used in the "resolvable condition". It can extract sharp images from diffraction-blurred images directly, and get full details of samples' spatial structure without using high frequency components. Therefore, it is termed "Deconvolutional Super-resolution (DeSu-re)". Popular convolution methods such as wiener filtering, usually requires images not to be cropped. But the proposed technique can work even without any observed image outside the ROI. In principle, it could be used for resolving the inner structures of luminous points even if they are infinitely small; or distinguishing multiple points at the same time, even if they overlap one another. Therefore, this technique could achieve unlimited resolutions in theory. When combined with existing techniques, it can "split" a single point into multiple smaller points and thus improves resolution by several times. The simulation experiments directly verify resolution improvement to 200%~300% (i.e., 2~3 times) for the spatial domain method, and 200%~400% (i.e., 2~4 times) for the frequency domain method. In addition, they also indirectly verify resolution improvement to 200%~10,000% (i.e., 2~100 times) for both methods. One of the future directions is to achieve higher resolution and verify the effectiveness in practice.

But there are still practical difficulties, especially the strong effect of the observed image's and PSF's distortion (e.g., noise) on the results. Therefore, it is also an important future direction to get practical results as close to simulation results as possible. With the development of imaging devices and the improvement of signal-noise ratio, the accuracy of the proposed technique would also improve accordingly. The proposed technique could be combined with other techniques such as conventional microscope, confocal microscope, existing super-resolution methods, and so on. By extracting more details directly from the data of these techniques, higher resolutions or efficiency could be achieved. For example, further resolve the inner details of individual molecules, fluorescent probes or tiny light sources after localization them. Or, several adjacent points (molecules, etc.) could be imaged and resolved at the same time, using the proposed technique.

## Acknowledgements

We thank THU, PKU, RUC, CAS, NLC, HDL, BISTU, HH and HL for their resources, information and support.